\documentstyle[12pt]{article}
\def\beq{\begin{equation}}
\def\eeq#1{\label{#1}\end{equation}}
\def\barr#1{\begin{equation}\begin{array}{#1}\displaystyle}
\def\earr#1{\end{array}\label{#1}\end{equation}}
\def\bit{\begin{itemize}}
\def\eit{\end{itemize}}
\def\ben{\begin{enumerate}}
\def\een{\end{enumerate}}
\def\bce{\begin{center}}
\def\ece{\end{center}}
\def\bmi{\begin{minipage}}
\def\emi{\end{minipage}}
\def\btab{\begin{tabular}}
\def\etab{\end{tabular}}

\def\dis{&\displaystyle}
\def\di#1{\\[0.#1cm]\displaystyle}
\def\r#1{(\ref{#1})}
\def\df{\stackrel{\rm def}{=}}
\def\ot#1#2{\textstyle \frac{#1}{#2}}
\def\od#1#2{\displaystyle \frac{#1}{#2}}
\def\lil#1{\hbox to\hsize{ #1 \hfil}}
\def\lilr#1#2{\hbox to\hsize{#1 \hfil #2 }}
\def\lir#1{\hbox to\hsize{\hfil #1 }}
\def\lic#1{\hbox to\hsize{\hfil #1 \hfil}}
\def\si{\sigma}
\def\ga{\gamma}

\def\al{\alpha}

\def\De{\Delta}

\def\Z{\indent \indent }

\def\six{\sigma ^{\xi }}
\def\sie{\sigma ^{\eta }}
\def\sio{\sigma ^{\xi \eta }}
\def\siu{\sigma ^{a}}
\begin{document}
\bce
{\Large \bf Electroweak Radiative Effects in Deep Inelastic\\
\vspace{2mm}  
Interaction of Polarized Leptons and Nucleons}\\
\vspace{4mm}  
{\large I. V. Akushevich, A.N. Ilyichev and N. M.Shumeiko}\\
\vspace{4mm}  
 {\it National Scientific and Education Center of Particle and
 High Energy Physics attached to Byelorussian State University}
\ece
\begin {abstract}
The results for one-loop correction to deep inelastic scattering of
longitu\-di\-nal polarized leptons on longitudinal polarized
hadrons are obtained
withing
the framework of the standard theory of electroweak interactions
and ordinary quark-parton model. The on-shell renormalization
scheme in  t'Hooft-Feyn\-man gauge is applied.

The numerical analysis is carried out under conditions  of
modern particle physics experiments.
Particular emphasis is laid on  contributions
 usually ignored at RC procedure --
electroweak corrections to electromagnetic asym\-me\-t\-ry and  RC
to hadronic current.
The structure  of RC contribution to polarized asymmetries within
the framework of  QED and electroweak theory is also
discussed.
\end {abstract}

\section{Introduction}
\Z Main sources of an information about spin properties of
nucleons are experiments on  deep inelastic scattering (DIS) of
polarized lepton by polarized targets
\cite{SLACold}-\cite{SLAC}.
Results of the last ones
 \cite{emc}-\cite{SLAC}
   form an inconsistent picture
 of  nature of nucleon spin (see \cite {JM} and reference therein).
This stimulates realization of new experiments \cite {HERA, FNAL} on
measurement of proton and neutron spin-dependent SF
$g_1^ {p, n} $ and $g_2^ {p, n} $ and on testing of
Bjorken \cite{Bjo}, Ellis-Jaffe \cite{EJ}  and
Burkhardt-Cottingham \cite{BG} sum rules.

An interpretation of experimental data requires an
adequate calculation of  background radiative effects  (RE).
The one-loop contribution  to cross section of DIS of
non-polarized
particles within the  framework of  quantum electrodynamics (QED)  was
widely discussed in literature (see, for example,
\cite{MoT}-\cite{10872}).
  Results of
 model-independent part of the one-loop QED contribution and
model-dependent RE in  hadronic current (hadronic RE) in DIS of polarized
particles can be found in \cite{KSh,TSh}.

    In the present report the
results for
one-loop radiative correction (RC) to DIS of polarized lepton on
polarized
proton are obtained
    within the framework of the standard theory of
electroweak interaction and ordinary parton model.
 We follow to  the on-mass renormalization scheme
in t'Hooft-Feynman gauge \cite{BHS} (see also the reviews
\cite{Aoki}-\cite{Den})
  and use the results of calculation for the
case of non-polarized particles \cite {BS, Bardin} \footnote {Some
estimates of the one-loop contribution in DIS
of polarized particles can be
found in article
\protect\cite{Kukhto}}.
We note, that the problem of infrared divergence cancellation
has been  resolved within covariant approach been
 proposed in \cite{BSh}: the RC
is to be finally free from dependence of
any unphysical parameter like "photon softness".

    In section \ref{Ma} we discuss one-loop RC to DIS of polarized
    particles.
It is shown, that RC can be
separated by leading-log contribution and contribution
independent of particle masses. In this report we give the explicit form
only for leading  log contribution.
In section \ref{Nu} infrared free one-loop RC
 is
studied numerically. We study the structure of contributions in
polarization asymmetry, calculate the magnitudes of effects normally
missed at RC procedure of experimental data and analyse
RC to SF $g_1(x,Q^2)$ and to sum rules.

\section{One-loop RC} \label{Ma}

\begin{figure}[p]
\vspace{-1cm}
\hspace{-1cm}
\begin{tabular}{cccc}
\begin{picture}(100,100)
\multiput(40,57)(0,8){3}{\oval(4.0,4.0)[r]}
\multiput(40,61)(0,8){3}{\oval(4.0,4.0)[l]}
\put(30,60){\line(2,-1){20.}}
\put(50,50){\line(2,1){20.}}
\put(30,10){\line(2,1){20.}}
\put(50,20){\line(2,-1){20.}}
\put(49,20){\line(0,2){30.}}
\put(51,20){\line(0,2){30.}}
\end{picture}
&
\begin{picture}(100,100)
\multiput(60,57)(0,8){3}{\oval(4.0,4.0)[l]}
\multiput(60,61)(0,8){3}{\oval(4.0,4.0)[r]}
\put(30,60){\line(2,-1){20.}}
\put(50,50){\line(2,1){20.}}
\put(30,10){\line(2,1){20.}}
\put(50,20){\line(2,-1){20.}}
\put(49,20){\line(0,2){30.}}
\put(51,20){\line(0,2){30.}}
\end{picture}
&
\begin{picture}(100,100)
\multiput(40,17)(0,8){3}{\oval(4.0,4.0)[r]}
\multiput(40,21)(0,8){3}{\oval(4.0,4.0)[l]}
\put(30,60){\line(2,-1){20.}}
\put(50,50){\line(2,1){20.}}
\put(30,10){\line(2,1){20.}}
\put(50,20){\line(2,-1){20.}}
\put(49,20){\line(0,2){30.}}
\put(51,20){\line(0,2){30.}}
\end{picture}
&
\begin{picture}(100,100)
\multiput(60,17)(0,8){3}{\oval(4.0,4.0)[l]}
\multiput(60,21)(0,8){3}{\oval(4.0,4.0)[r]}
\put(30,60){\line(2,-1){20.}}
\put(50,50){\line(2,1){20.}}
\put(30,10){\line(2,1){20.}}
\put(50,20){\line(2,-1){20.}}
\put(49,20){\line(0,2){30.}}
\put(51,20){\line(0,2){30.}}
\end{picture}
\\[1cm]
\begin{picture}(100,100)
\put(30,60){\line(2,-1){20.}}
\put(50,50){\line(2,1){20.}}
\put(30,10){\line(2,1){20.}}
\put(50,20){\line(2,-1){20.}}
\multiput(50,48)(0,-8){2}{\oval(4.0,4.0)[r]}
\multiput(50,44)(0,-8){2}{\oval(4.0,4.0)[l]}
\multiput(50,26)(0,8){2}{\oval(4.0,4.0)[r]}
\multiput(50,22)(0,8){2}{\oval(4.0,4.0)[l]}
\put(50,35){\circle*{10.}}
\put(42,44){\makebox(0,0){\footnotesize $\gamma $}}
\put(42,26){\makebox(0,0){\footnotesize $\gamma $}}
\end{picture}
&
\begin{picture}(100,100)
\put(30,60){\line(2,-1){20.}}
\put(50,50){\line(2,1){20.}}
\put(30,10){\line(2,1){20.}}
\put(50,20){\line(2,-1){20.}}
\multiput(50,48)(0,-8){2}{\oval(4.0,4.0)[r]}
\multiput(50,44)(0,-8){2}{\oval(4.0,4.0)[l]}
\multiput(50,26)(0,8){2}{\oval(4.0,4.0)[r]}
\multiput(50,22)(0,8){2}{\oval(4.0,4.0)[l]}
\put(50,35){\circle*{10.}}
\put(42,44){\makebox(0,0){\footnotesize $\gamma $}}
\put(42,26){\makebox(0,0){\footnotesize $Z $}}
\end{picture}
&
\begin{picture}(100,100)
\put(30,60){\line(2,-1){20.}}
\put(50,50){\line(2,1){20.}}
\put(30,10){\line(2,1){20.}}
\put(50,20){\line(2,-1){20.}}
\multiput(50,48)(0,-8){2}{\oval(4.0,4.0)[r]}
\multiput(50,44)(0,-8){2}{\oval(4.0,4.0)[l]}
\multiput(50,26)(0,8){2}{\oval(4.0,4.0)[r]}
\multiput(50,22)(0,8){2}{\oval(4.0,4.0)[l]}
\put(50,35){\circle*{10.}}
\put(42,44){\makebox(0,0){\footnotesize $Z $}}
\put(42,26){\makebox(0,0){\footnotesize $\gamma $}}
\end{picture}
&
\begin{picture}(100,100)
\put(30,60){\line(2,-1){20.}}
\put(50,50){\line(2,1){20.}}
\put(30,10){\line(2,1){20.}}
\put(50,20){\line(2,-1){20.}}
\multiput(50,48)(0,-8){2}{\oval(4.0,4.0)[r]}
\multiput(50,44)(0,-8){2}{\oval(4.0,4.0)[l]}
\multiput(50,26)(0,8){2}{\oval(4.0,4.0)[r]}
\multiput(50,22)(0,8){2}{\oval(4.0,4.0)[l]}
\put(50,35){\circle*{10.}}
\put(42,44){\makebox(0,0){\footnotesize $Z $}}
\put(42,26){\makebox(0,0){\footnotesize $Z $}}
\end{picture}
\\[1cm]
\begin{picture}(100,100)
\put(30,60){\line(2,-1){20.}}
\put(50,50){\line(2,1){20.}}
\put(30,10){\line(2,1){20.}}
\put(50,20){\line(2,-1){20.}}
\put(49,20){\line(0,2){30.}}
\put(51,20){\line(0,2){30.}}
\put(38,56){\line(2,0){24.}}
\put(42,54){\line(2,0){16.}}
\end{picture}
&
\begin{picture}(100,100)
\put(30,60){\line(2,-1){20.}}
\put(50,50){\line(2,1){20.}}
\put(30,10){\line(2,1){20.}}
\put(50,20){\line(2,-1){20.}}
\put(49,20){\line(0,2){30.}}
\put(51,20){\line(0,2){30.}}
\put(38,14){\line(2,0){24.}}
\put(42,16){\line(2,0){16.}}
\end{picture}
&
\begin{picture}(100,100)
\put(30,60){\line(2,-1){10.}}
\put(60,55){\line(2,1){10.}}
\put(30,10){\line(2,1){20.}}
\put(50,20){\line(2,-1){20.}}
\put(40,55){\line(2,0){20.}}
\put(49,20){\line(0,2){25.}}
\put(51,20){\line(0,2){25.}}
\multiput(42,55)(4,-4){3}{\oval(4.0,4.0)[lb]}
\multiput(42,51)(4,-4){2}{\oval(4.0,4.0)[tr]}
\multiput(50,47)(4,4){3}{\oval(4.0,4.0)[br]}
\multiput(54,47)(4,4){2}{\oval(4.0,4.0)[tl]}
\put(40,44){\makebox(0,0){\footnotesize $W$}}
\put(60,44){\makebox(0,0){\footnotesize $W$}}
\end{picture}
&
\begin{picture}(100,100)
\put(30,10){\line(2,1){10.}}
\put(60,15){\line(2,-1){10.}}
\put(30,60){\line(2,-1){20.}}
\put(50,50){\line(2,1){20.}}
\put(40,15){\line(2,0){20.}}
\put(49,50){\line(0,-2){25.}}
\put(51,50){\line(0,-2){25.}}
\multiput(42,15)(4,4){3}{\oval(4.0,4.0)[lt]}
\multiput(42,19)(4,4){2}{\oval(4.0,4.0)[br]}
\multiput(50,23)(4,-4){3}{\oval(4.0,4.0)[tr]}
\multiput(54,23)(4,-4){2}{\oval(4.0,4.0)[bl]}
\put(40,25){\makebox(0,0){\footnotesize $W$}}
\put(60,25){\makebox(0,0){\footnotesize $W$}}
\end{picture}
\\[1cm]
\begin{picture}(100,100)
\put(30,60){\line(2,-3){10.}}
\put(40,45){\line(2,0){20.}}
\put(60,45){\line(2,3){10.}}
\put(30,10){\line(2,3){10.}}
\put(40,25){\line(2,0){20.}}
\put(60,25){\line(2,-3){10.}}
\put(40,25){\line(0,2){20.}}
\put(42,25){\line(0,2){20.}}
\put(58,25){\line(0,2){20.}}
\put(60,25){\line(0,2){20.}}
\end{picture}
&
\begin{picture}(100,100)
\put(30,60){\line(2,-3){10.}}
\put(40,45){\line(2,0){20.}}
\put(60,45){\line(2,3){10.}}
\put(30,10){\line(2,3){10.}}
\put(40,25){\line(2,0){20.}}
\put(60,25){\line(2,-3){10.}}
\put(40,26){\line(1,1){19.}}
\put(41,25){\line(1,1){19.}}
\put(40,44){\line(1,-1){19.}}
\put(41,45){\line(1,-1){19.}}
\end{picture}
&
\begin{picture}(100,100)
\put(30,60){\line(2,-3){10.}}
\put(40,45){\line(2,0){20.}}
\put(60,45){\line(2,3){10.}}
\put(30,10){\line(2,3){10.}}
\put(40,25){\line(2,0){20.}}
\put(60,25){\line(2,-3){10.}}
\multiput(40,27)(0,8){3}{\oval(4.0,4.0)[l]}
\multiput(40,31)(0,8){2}{\oval(4.0,4.0)[r]}
\multiput(60,27)(0,8){3}{\oval(4.0,4.0)[r]}
\multiput(60,31)(0,8){2}{\oval(4.0,4.0)[l]}
\put(33,35){\makebox(0,0){\footnotesize $W$}}
\put(67,35){\makebox(0,0){\footnotesize $W$}}
\end{picture}
&
\begin{picture}(100,100)
\put(30,60){\line(2,-3){10.}}
\put(40,45){\line(2,0){20.}}
\put(60,45){\line(2,3){10.}}
\put(30,10){\line(2,3){10.}}
\put(40,25){\line(2,0){20.}}
\put(60,25){\line(2,-3){10.}}
\multiput(42,25)(4,4){5}{\oval(4.0,4.0)[lt]}
\multiput(42,29)(4,4){5}{\oval(4.0,4.0)[br]}
\multiput(42,45)(4,-4){5}{\oval(4.0,4.0)[lb]}
\multiput(42,41)(4,-4){5}{\oval(4.0,4.0)[tr]}
\put(40,35){\makebox(0,0){\footnotesize $W$}}
\put(60,35){\makebox(0,0){\footnotesize $W$}}
\end{picture}
\end{tabular}
\caption{A complete set of electroweak graphs,
contributed to lepton-quark scattering within quark-parton model. The
double line corresponds to the contribution of $ \gamma$- or
$Z$-exchange.
 All possible graphs, which give the contribution to
vacuum polarization,
are designated by symbol $ \bullet $
\protect\cite{Holl}.
}
\label{Fg}
\end{figure}
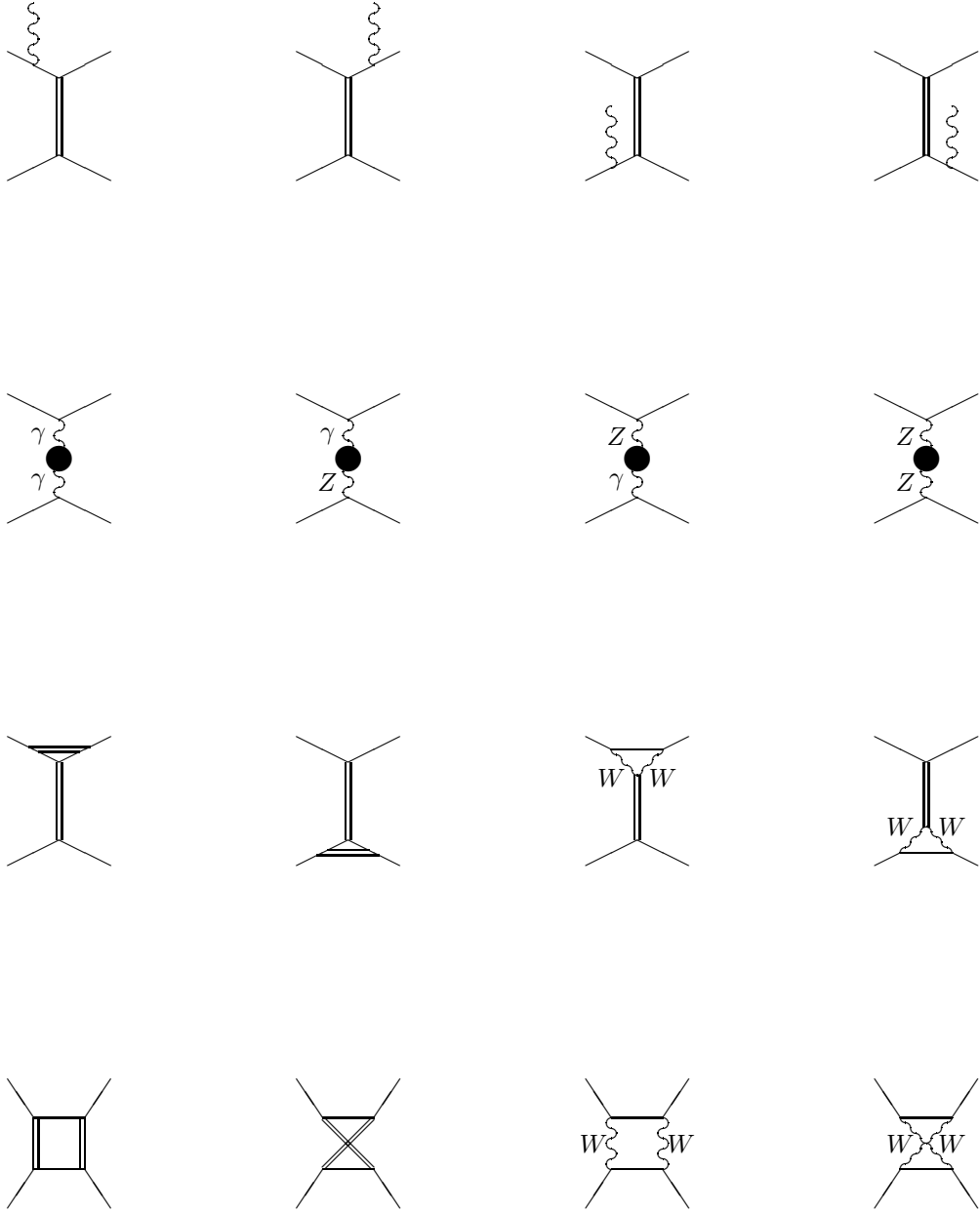

\Z
 We consider DIS of longitudinally polarized leptons on
longitudinally polarized nucleons
\beq
\ell (k_ {1}, \xi) + N (p, \eta) \rightarrow \ell (k_ {2}) + X
\eeq{proc}
with taking into account the
one-loop RE within parton model.
The vectors in brackets designate a momentum and polarization of
appropriate
particles ($k_1^2=k_2^2=m^2$, $p^2=M^2$). A complete set of Feynman
graphs is presented in figure \ref{Fg}. The result can be written in
 a form of  sum of the
Born contribution, R- and V-contributions (the first line of
graphs in figure \ref{Fg} and the last ones)
\begin{equation}
  {\sigma} = {\sigma _ {0}} + {\sigma _V} + {\sigma _R},
\end{equation}
    where $ \sigma_ {0, V, R} \df d^2 \sigma_ {0, V, R} / dxdy$, and
    $x$, $y$ are scaling variables.

The on-shell renormalization scheme of
electroweak theory
is
submitted in the reviews \cite {BHS, Aoki} and is advanced in
\cite{Holl}-\cite{Den}.
 This scheme uses an electrical charge and masses of particles as physical
parameters.
 Both
 self-energies together with a complete set of
renormalization
constants and  vertex functions of fermions were calculated
 in \cite {BHS}.
 Renormalization formulae for
self-energies of vector bosons, vertex functions and graphs of
  two boson exchange were also presented. All results of  \cite{BHS}
   are given in a
form convenient for further applications to RC calculation for fermion
processes,
including DIS. We use these results for calculation of  V-contribution.

V-contribution can be written  as a sum of the contributions
of self-energies of vector boson and quark ($ \sigma^B_S$, $
\sigma^f_S$),
vertex functions of lepton and hadron ($ \sigma_ {Vl} $, $ \sigma_ {Vq} $)
and contributions of two boson exchange ($ \sigma_ {box} $):
\begin{equation}
\begin{array}{l}
\displaystyle
\sigma_ {V} =
\sigma^B_S+\sigma^f_S+
\sigma_{Vl}
+\sigma_{Vq}+\sigma_{box}.
\end{array}
\label{crs}
\end{equation}

As usual we keep
the leading contributions (containing mass singularity
--- $\ln m^2$) and
next-to-leading ones (without any mass dependence). In this report we
give the explicit form only for leading contributions (for more details
see \cite{AISh}).

    The mass singularity in the V-contribution due to smallness of
mass of particles participating in a scattering, is contained
 only in the correction to vertex:
\begin{equation}
\begin{array}{l}
\displaystyle
{\sigma _ {Vl}} = \frac {\alpha} {2 \pi} \Lambda_1 (Q^2, m^2) {\sigma _
{0}},
\\[0.5 cm] \displaystyle {\sigma _ {Vq}} = \frac {\alpha} {2 \pi}
\sum_q e_q^2 \Lambda_1 (Q^2, m_q^2) {\sigma _0^q}.
\end{array}
\label{vsin}
\end{equation}
Here $e_q$ is quark charge, and $\displaystyle
\si_0=\sum_q\si_0^q=\sum_{ij=\ga
Z}\si^{ij}$. The function $ \Lambda_1 (Q^2, m^2) $ is given by
the formula (B.3) of \cite {BHS}. In our case it has the form
\beq
\Lambda_1(Q^2,m^2)
= \ln \frac {Q^2} {m^2} \left (1 + \ln \frac {Q^2} {m^2} \right).
\eeq{g001}

Self-energies of vector bosons
\begin{equation}
\sigma^B_S=
-2\Pi^{\gamma}\sigma_0^{\gamma\gamma}
-2(\Pi^{\gamma }+\Pi^Z)\left(
\sigma_0^{\gamma Z}+\sigma_0^{Z \gamma}
\right) -2\Pi^Z\sigma_0^{ZZ}
-2\Pi^{\gamma Z}\sigma^{\gamma Z}_S
\label{se}
\end{equation}
have mass singularity due to smallness
of particle mass appearing in loops of vacuum polarization.
The quantities $\Pi ^{\ga,Z,\ga Z}$ are polarization operators and
$\si_S^{\ga Z}$ was found in \cite{AISh}. For simplicity we give
expressions for vacuum polarization by leptons only.
For the various polarization operators (\ref{se})
we have
\begin{equation}
\begin{array}{l}
\displaystyle
\Pi ^ {\gamma} =- \frac {\alpha} {3 \pi} \sum_ {l=e, \mu, \tau} (v_l^
{\gamma}) ^2 \ln \frac {Q^2} {m_l^2},
\\[0.5 cm] \displaystyle \Pi ^ {\gamma Z} =- \frac {\alpha} {3 \pi}
\sum_ {l=e, \mu, \tau}
\biggl(v_l^{\gamma
}v_l^Z-\frac{c_w}{s_w}
\Bigl ((v_l^ {Z}) ^2 + 3 (a_l^ {Z}) ^2- (v^ {W} _l) ^2- (a^ {W} _l) ^2
\Bigr)
\biggr)
\ln \frac {Q^2} {m_l^2}, \\[0.5 cm] \displaystyle \Pi ^ {Z} =- \frac
{\alpha} {3 \pi} \sum_ {l=e, \mu, \tau}
\biggl(\frac{c_w^2}{s_w^2}
\Bigl((v_l^{Z})^2+3(a_l^{Z})^2\Bigr)
-\frac{c_w^2-s_w^2}{s_w^2}
\Bigl((v^{W}_l)^2+(a^{W}_l)^2\Bigr)
\biggr)
\ln \frac {Q^2} {m_l^2}
,
\end{array}
\label{Piur}
\end{equation}
where $v_l$ and $a_l$ are vector and axial coupling constants of leptons,
and $c_w$ and $s_w$ are cosine and sine of Weinberg angle.

Result for the contribution of process with radiation of real photon
\begin{equation}
\ell (k_ {1}, \xi) + р (p, \eta) \rightarrow \ell (k_ {2}) + \gamma (k)
+ X
\end{equation}
to observed cross  section of DIS can be written in the form
\beq
{\sigma _R} = {\alpha \over 2 \pi} \ln \frac {Q^2} {\lambda^2} \sum_q
J (Q^2,0)
{\sigma_0^q}
+ {\alpha \over \pi} \sum_q \delta_q {\sigma_0^q} + \sum_q {\sigma
_R^q},
\eeq{rv}
where
\beq
\sigma _R^q=
\sum_{ij=\gamma ,Z} \left\{
\sigma ^{ij}_l
+{\hat \sigma }^{ij}_l
+e_q \sigma ^{ij}_{lh}
+e_q^2\left(\sigma ^{ij}_h
+{\hat \sigma }^{ij}_h
\right)
\right\}.
\eeq{sigrad}
Low index
($b=l,h,lh$) of cross sections in right side of the equation
corresponds to contributions of radiation by leptons, by hadrons and
their interference. Only $\sigma ^{ij}_l$ and $\sigma ^{ij}_h$
content leading contributions.

The infrared divergence is extracted
 by the method of Bardin and Shumeiko \cite{BSh}.
It is completely contained in the first term of expression (\ref {rv})
and
is cancelled with an appropriate term of V-contribution.

 We
distinguish three  kinds of mass singularities:
leptonic ($ \ln \; Q^2/ m^2$), quark ($ \ln \; Q^2/ m_q^2$)
and nucleon ($
\ln \; Q^2/ M^2$). Such singularities are contained
in quantity $\delta_q $:
\beq
\delta_q= -\ot12l_m^2 +l_m(2l_v+l_{sx}+1)
+e_q^2\bigl(\l_q(2l_v-\ot14)\bigr)
,
\eeq{g002}
where
\beq
\l_m=\ln\frac{Q^2}{m^2},     \;\;\;
\l_q=\ln\frac{Q^2}{m_q^2},     \;\;\;
\l_v=\ln\frac{1-x}{x},   \;\;\;
\l_{sx}=\ln\frac{y^2}{1-y},
\eeq{g030}
and in $ {\sigma^{ij}_{l, h}} $, which are considered below.

Firstly we consider  dependence on leptonic mass $m$,
which is
 contained only in the contribution of radiation by leptons
 $\sigma^ {ij} _l$.
Result in standard leading log form is obtained by
splitting the cross section by the contributions appearing from
$k_1$- and
$k_2$-peaks \cite{Sh}:
\begin{equation}
\sum_{ij} \sum_q {\sigma ^{ij} _l} =
{\sigma ^{k_1} _R} + {\sigma ^{k_2} _R},
\end{equation}
and
\begin{equation}
\begin{array}{l}
\displaystyle
{\sigma ^{k_1} _R} = \frac {\alpha} {2 \pi} l_m
\int\limits_{z_1^l}^1
{ydz_1 \over z_1-1 + y} \left \{{1 + z_1^2 \over 1-z_1} {\sigma ^ {k_1}
_0} - {2 \over 1-z_1} {\sigma _0} \right \},
\\[0.5 cm] \displaystyle {\sigma ^ {k_2} _R} = \frac {\alpha} {2 \pi}
l_m
\int\limits_{z_2^l}^1
{ydz_2 \over z_2-1 + y} \left \{
{1 + z_2^2 \over z_2 (1-z_2)} {\sigma ^ {k_2} _0} - {2 \over 1-z_2} {
\sigma _0} \right \}.
\end{array}
\end{equation}
 Here the cross sections $ \sigma^{k_{1,2}}_0$ are obtained from born
one
\beq
{\sigma _0} \equiv {\sigma _0} \bigl (S, x, y \bigr)
\eeq{g013}
by replacements
\beq
{\sigma ^ {k_1} _0} = {\sigma _0} \biggl (z_1 S, \od{xyz_1}{z_1-1+y},
\od{z_1-1+y}{z_1}
\biggr),
\qquad
{\sigma ^ {k_2} _0} = {\sigma _0} \biggl (S, \od{xy}{z_2-1+y},
\od{z_2-1+y}{z_2}
\biggr).
\eeq{g005}
The low limits of integration are equal to
\begin{equation}
z_1^l=(1-y)/(1-xy),\;\;
z_2^l=1-y+xy.
\end{equation}

 The dependence on quark mass is retained only in the
contributions of
radiation by hadrons and in the leptonic QED correction.
 For the contribution of hadronic radiation
$\sigma^{ij}_h$
we obtain
\begin{equation}
\sum_q e_q^2 {\sigma _h^ {ij}} = {\sigma _0^ {ij}} \left (f^ {\pm} _q
(x) \rightarrow f_q^ {\pm rad} (x) \right),
\end{equation}
where
\begin{equation}
\begin{array}{l}
\displaystyle
f_q^{\pm rad} (x) = e_q^2 \frac {\alpha} {2 \pi} l_q
\biggl\{
-f_q^{\pm}(x)
\bigl(l_q+2l_v\bigr)
\di5
 \qquad \qquad + \int \limits ^ {1} _ {x} \od {dz} {z} \biggl \{
{1 + z^2 \over 1-z} f_q^ {\pm} (x / z) - {2 \over 1-z} f_q^ {\pm} (x)
\biggr\}
\biggr\}.
\end{array}
\label{fff}
\end{equation}

  In the case of leptonic electromagnetic radiation the appearing
of quark mass has a
purely kinematic origin, and it can be replaced with proton mass
in according to the rule of parton model:
$m_q= \xi M$. It follows from comparison of results for electromagnetic
radiation in parton model and ones obtained by a model
independent way. In this case we have
\begin{equation}
{{\sigma} ^ {\gamma \gamma} _ {l}} = {\alpha ^3 y \over 4} \ln \frac
{Q^2} {M^2} \int \limits_x^1 {d \xi \over \xi} \left \{T_ {+ M} R^ {
\ga \ga} _VF^ {q \ga \ga} _V (\xi) + T_ {-M} R^ {\ga \ga} _AF^ {q \ga
\ga} _A (\xi)
\right\},
\end{equation}
where
\begin{equation}
T_ {\pm M} =- {1\pm(1-y)^2 \over Sy(1-y)} {1\pm(1-x/\xi)^2 \over
x^2}.
\end{equation}

Thus, self-energies, vertex functions and the contributions of
radiation by
leptons and by hadrons have a mass singularity and therefore are
significant.
Neither the lepton-hadron interference in bremsstrahlung nor
the contributions of
two boson exchange contain any mass singularities. In this
sense we
speak, that the leptonic and hadronic corrections are separated.
Radiation by leptons contains both leptonic and nucleon mass
singularity.
Extraction of contribution containing the leptonic mass in a
separate term leads to
peaking and
leading log  approximation.
The contribution of a nucleon mass singularity
corresponds to t-peak when a real photon is radiated parallel to
virtual one.
 This contribution is not extracted by methods of the leading
logarithms. The quark mass singularity in purely hadronic radiation
is reduced to the correction to parton distributions (\ref {fff}). In
 the leading log approximation this result for
unpolarization and polarization cases was received in refs. \cite
{ll, AK}.

\section{Numerical analysis}\label{Nu}

\Z
In this section complete one-loop RC to various observables in
DIS of
polarized particles is analyzed numerically in a wide range kinematic
variables. The special attention is paid to the contributions,
which are usually
neglected at the data analysis in modern polarization
experiments: to the
electroweak corrections to electromagnetic asymmetry and
effects of radiation by hadrons.
Important question on the structure of the contributions in
polarization
asymmetry in QED and  electroweak theory is discussed
as well.

\subsection{Structure of contribution to polarization asymmetry}

\Z The cross section of DIS $ \si $ both at a born level and at
a level of the radiative corrections
 can be written as follows:
\beq
\si = \siu + P_L \six + P_N \sie + P_LP_N \sio.
\eeq{g007}
Here $ \siu$ is unpolarization cross section, and three other terms
give
polarization contributions. $P_L$ and $P_N$ are polarization degrees of
lepton and hadron.

Let us define the following quantities:
\beq
A^{\xi} = {\six \over \siu}, \quad A^{\eta} = {\sie \over \siu},
\quad A^{\xi \eta} = {\sio \over \siu}
\eeq{AAA}
and consider polarization asymmetry
\begin{equation}
A= {\sigma^{\uparrow \uparrow} - \sigma^{\uparrow \downarrow} \over
\sigma^ {\uparrow \uparrow} + \sigma^ {\uparrow \downarrow}} =
{A^{\eta} + A^{\xi \eta} \over 1 + A^{\xi}}.
\label{As}
\end{equation}
We note, that $A$ equals to $A^{\xi \eta} $ in electrodynamics. By
using as an example $A^{\xi \eta} $ we analyse magnitudes of the various
one-loop contributions in  asymmetry.
Such analysis is convenient to conduct by expansion
$A^{\xi \eta} $ with
the account RC in a series over coupling constants:
\beq
A^{\xi \eta} = {\sio _0 + \sum_i \sio _i \over \siu _0 + \sum_i
\siu _i} = A^ {\xi \eta} _0 + \frac {1} {\sigma _0^ {a^2}} \sum_i \bigl
(\sio _i \siu _0 - \siu _i \sio _0 \bigl) \; + O (\alpha^2),
\eeq{0013}
where quantities with an index "0" are the born contributions.
The sum on
$i$ corresponds to separating of the one-loop correction into the
contributions:
   effects of vacuum polarization,
   corrections to vertex, two boson
exchange
 and bremsstrahlung of photon (\ref {rv}).
>From (\ref {0013}) we see  if for any cross section
$ \sigma_i^ {a,\xi \eta} $ a quantity
\beq
 \sigma _i^{\xi \eta} \sigma _0^ {a} - \sigma _i^ {a} \sigma _0^ {\xi
 \eta}
\eeq{kre}
is equal to zero, then it does not give the contribution to
polarization asymmetry.

In QED the spin average and spin dependent parts of one-loop
cross section can be written as
\beq
\sigma ^ {a, \xi \eta} = \sigma _0^ {a, \xi \eta}
+ \bigl (-2 \Pi_ {\gamma} ^ {\rm QED} + \delta F^ {\rm QED} _V +
\delta^ {a, \xi \eta} _ {2 \gamma} \bigr) \sigma _0^ {a, \xi \eta} +
\sigma _R^ {a, \xi \eta}.
\eeq{g009}
The quantities $ \delta^ {a, \xi \eta} _ {2 \gamma} $ describe the
contribution of two-photon exchange and are agreed
 with ones
considered in \cite {10872, TSh}. A symbol "QED"
indicates
 that only QED effects are retained in corresponding quantities.
 Expansion (\ref {0013}) in this
case gives
\beq
A=A_0 \bigl (1-
\delta^ {a} _ {2 \gamma} + \delta^ {\xi \eta} _ {2 \gamma} \bigr)
+ \bigl (\sigma _R^ {\xi \eta} \sigma _0^ {a}
- \sigma _R^ {a} \sigma _0^ {\xi \eta} \bigr) / {\sigma _0^ {a^2}} \; +
O (\alpha^2).
\eeq{0055}
Among all terms of V-contribution only small (without mass singularity
terms) effects of two-photon exchange contribute to
polarization asymmetry.
R-contributi\-on has a leading log term and dominates in
 (\ref{0055}).
The logarithmic correction to polarization
 asymmetry has been calculated in \cite {AK}.

In the electroweak theory the contributions of vertex functions
 and self-energies are not
factorized in front of born section and
do not vanish in combination (\ref {kre}).
That is valid for leading log contributions of polarization operators
\r{Piur} and next-to-leading terms of vertex functions.
 Thus, in contrast to QED,
where only bremsstrahlung contributes to polarization asymmetry, in
electroweak theory
V-contribution is also significant.

\subsection{Electroweak effects and radiation by hadrons}

\Z Cross section of DIS with taking into account
 the complete one-loop correction can be
presented by
\beq
\si=\sum_b
\sum_B \si^ {bB}.
\eeq{0040}
The index $B= \ga, Z, I$ corresponds to the contributions $ \ga $-,
$Z$-exchange and their interference, and $b=0, l, h, i$ --- to born
contribution,
leptonic, hadronic correction and lepton - hadronic interference.
We note, that in modern polarization experiments the procedure
RC takes
into account only contribution $ \si ^ {l \ga} $, and the systematic error
due to RC
includes only an error of calculation of the leptonic
QED correction.

Below we consider the electroweak effects at born and one-loop
levels
 and contributions of radiation by hadrons.
For convenience of the numerical analysis we define the next
cross sections:
\barr{ll}
\si^0= \sum_ {\scriptscriptstyle B= \ga, I, Z}
\si^{0B},
\dis
\si^{had}=
\si^{0\ga}+
\sum_ {\scriptscriptstyle b=l, i, h}
\si^{b\ga},
\di5
\si^{lep}_z=
\si^{0\ga}+
\sum_ {\scriptscriptstyle B= \ga, I, Z}
\si^{lB},
\dis
\si^{had}_z=
\si^{0\ga}+
\sum_ {\scriptscriptstyle b=l, i, h}
\sum_ {\scriptscriptstyle B= \ga, I, Z}
\si^{bB},
\earr{0045}
 In each of them in addition to the contributions $ \si ^ {l \ga} $
 and/or $ \si^{0 \ga} $ one of the following effects are added:
QED radiation by hadrons ($ \si ^ {had} $) and
$Z$-exchange: at a born level ($ \si ^ {0} $), at bremsstrahlung by
lepton ($ \si ^
{lep} _z$) and hadron ($ \si ^ {had} _z$).
Using \r {0045}, we construct of asymmetry \r {As} --- $A_0$, $A^
{had} $, $A^ {lep} _z$ and $A^ {had} _z$.
 In the range of small $x$ the basis contribution
at
an one-loop level is given by the QED correction, including correction
to
hadronic current.
The contribution of the graphs with $Z$-exchange is
insignificant  both at a
born level ($A^ {\ga \ga} _0 \approx A_0$) and on a level of RC
($A^ {\ga \ga} \approx A_z^ {lep} $,
$A \approx A^ {had} \approx A_z^ {had} $). In the range of high $x$
 electroweak effects dominate
and the complete correction to asymmetry
is defined basically by an electroweak interference ($A \approx
 A_0$).
The influence of RE decreases with growth  $x$: so, for example,
the part of RE becomes smaller 10\% for $E_1=100$ГэВ at
$x \geq 0.8$, and for $E_1=1000$GeV already at $x \geq 0.15$.

We also  study numerically the next relative
corrections:
\begin{equation}
\begin{array}{ll}
\displaystyle
\Delta _B= {A_ {0} - A_ {0} ^ {\gamma \gamma} \over A_ {0} ^ {\gamma
\gamma}},
 \dis
\Delta _ {h} = {A^ {had}
- A^ {\gamma \gamma} \over A^ {\gamma \gamma}}, \\[0.5 cm]
\displaystyle \Delta _ {ew} = {A^ {lep} _z
- A^ {\gamma \gamma} \over A^ {\gamma \gamma}},
 \dis
\Delta _ {hz} = {A_ {z} ^ {had}
- A^ {\gamma \gamma} \over A^ {\gamma \gamma}}.
\end{array}
\label{0014}
\end{equation}
 $ \Delta _B$ is constructed from born asymmetries and gives the
correction
due to electroweak interference to  purely
electromagnetic born contribution. The other quantities
are investiga\-ted in
comparison with an one-loop model independent part of correction.
Thus, the corrections (\ref {0014}) (tab. \ref{tab4}) give insight on
values
of effects, not included by the usual QED procedure RC.
 The table also illustrates a dependence of discussed quantities
on
energy of scattering lepton. We note, that if at existing energies
of the correction (\ref{0014}) does not exceed 3-5 \%,
these effects can not be  ignored in future experiments with
 energies up to 1 TeV.

\begin{table}[t]
\begin{tabular}{||cc|cccc|cccc||}
\hline
$x$&$y$&
\multicolumn{4}{|c}{$E_1={\rm 100}\; GeV$}&
\multicolumn{4}{|c||}{$E_1={\rm 1000}\; GeV$} \\
\cline{3-{10}}
&& $\Delta _ {B}$& $\Delta _ {ew}$& $\Delta _{h}$&
$\Delta _ {hz}$& $\Delta _ {B}$& $\Delta _ {ew}$& $\Delta _ {h}$&$\Delta _ {hz}$\\
\hline
0.015&0.604&0.011&0.007&0.786& 0.792& 0.110& 0.103& 0.950&1.050\\
0.050&0.394&0.057&0.017&0.665& 0.681& 0.565& 0.228& 0.792&1.021\\
0.241&0.199&0.405&0.043&0.385& 0.422& 4.021& 0.473& 0.464&0.932\\
0.470&0.176&0.881&0.037&0.043& 0.061& 8.670& 0.241&0.046&0.129\\
0.950&0.168&1.847&-0.163&-0.010& -0.322& 17.893& -2.787& -0.029& -4.695\\
\hline
\end{tabular}
\caption{
 Corrections
$\Delta _{B}$, $\Delta _{ew}$,
$\Delta _{h} $, $\Delta _{hz}$
 in SMC kinematics \protect\cite{CERN}.}
\label{tab4}
\end{table}

\subsection{QED correction to $g_1(x,Q^2)$ and sum rules}

\Z RC procedure based on model independent exact formulae for the
lowest order RE
and described in \cite{ASh} gives as a result SF $g_1(x,Q^2)$
with taking into account only model independent effects. Model
dependent effects (electroweak effects, radiation by hadrons) are
ignored. In this case the effects should be taking into account for
subsequent analysing of the SF in parton model:
\begin{equation}
  g_1(x,Q^2) =g_1^0(x)+g_1^{QCD}(x,Q^2)+g_1^{QED}(x,Q^2),
\end{equation}
where
\beq
  g_1^0(x) =\od12 \sum_q e_q^2 f^{(-)}_q(x),
\eeq{g020}
and $g_1^{QCD}(x,Q^2)$ and $g_1^{QED}(x,Q^2)$ are QCD and QED correction
to it. In the report we discuss the QED effects which arise from
real photon radiation by hadrons and quark vertex function.
Analogously we obtain for each quark flavour $j$
\begin{equation}
 \De q_j =\De q_j^0+\De q_j^{QCD}+\De q_j^{QED},
\end{equation}
where each $\De q_j=\int_0^1dx(f^{(-)}_j(x)+\bar f^{(-)}_j(x))$.
We have to take into account the RC to
the observables $g_1(x,Q^2)$, $\De q_j$
and to obtain
the quantities $g_1^0(x)$,
$\De q_j^0$ as a experimental results.
By consideration of leading and next-to-leading contribution we have for
$g_1^{QED}(x,Q^2)$
\barr{l}
  g_1^{QED}(x,Q^2) = \od{\al}{4\pi}
  \sum_q e_q^4  \biggl\{
\left(\ot32 l_q +2l_ql_v-l_v^2-\ot72l_v-\ot52-\pi^2/3\right)
f^{(-)}_q(x)
\di5   \qquad  \qquad  \qquad
+ \int\limits_x^1 \od{dz}{z(1-z)}
\Bigl[
\left((1+z^2)(l_q-\ln z(1-z)-4)+5z-\ot12\right)
 f^{(-)}_q(\od x z)
\di5   \qquad \qquad   \qquad  \qquad \qquad  \qquad
-2
\left(l_q+\ln \od{z}{1-z}-\ot74\right)
   f^{(-)}_q(x)
\Bigr]
\biggr\}
\earr{g021}
and for $\De q_j^{QED}$
\beq
\De q_j^{QED}=-\od{9e_q^2\al}{4\pi}
\De q_j^0.
\eeq{g022}
The correction does not contain a leading contribution and is small. By
applying this result to
 EMC
results \cite{emc}
\beq
\De u= 0.782, \qquad
\De d=-0.472, \qquad
\De s=-0.190
\eeq{g024}
obtained for $Q^2=10.7{\rm GeV}^2$ and recalculated
with
taking into account
$\De q_j^{QCD}$ we find
\beq
\De u^0= 0.780, \qquad
\De d^0=-0.472, \qquad
\De s^0=-0.190.
\eeq{g025}

\end{document}